\documentclass{aa}
\usepackage{graphicx}
\usepackage{epsfig}
\usepackage{graphics}
\begin{document}%
   \title{Non-thermal processes in bowshocks of runaway stars}
\subtitle{Application to $\zeta$ Oph}

   \author{M.~V. del Valle\inst{1,2,}\thanks{Fellow of CONICET, Argentina}
           \and G.~E. Romero\inst{1,2,}\thanks{Member of CONICET, Argentina}
          }

   \offprints{Mar\'{\i}a V. del Valle : \\ {\em maria@iar-conicet.gov.ar}}
   \titlerunning{Non-thermal processes in bowshocks of runaway stars}

\authorrunning{del Valle \& Romero}  \institute{Instituto Argentino de
Radioastronom\'{\i}a, C.C.5, (1894) Villa Elisa, Buenos Aires,
Argentina \and Facultad de Ciencias Astron\'omicas y Geof\'{\i}sicas,
Universidad Nacional de La Plata, Paseo del Bosque, 1900 La Plata, Argentina}

\date{Received / Accepted}


\abstract 
{Runaway massive stars are O- and B-type stars with high spatial
velocities with respect to the interstellar medium. These stars can produce bowshocks in the surrounding gas. Bowshocks develop as arc-shaped structures, with bows pointing to the same direction as the stellar velocity, while the star moves supersonically through the interstellar gas. The piled-up shocked matter  emits thermal radiation and a population of locally accelerated relativistic particles is expected to produce non-thermal emission over a wide range of energies.} 
{We aim to model the non-thermal radiation produced in these sources.} 
{Under some assumptions, we computed the non-thermal emission produced by the relativistic particles and the thermal radiation caused by  free-free interactions, for  O4I and  O9I stars. We applied our model to $\zeta$ Oph (HD 149757), an intensively studied massive star seen from the northern hemisphere. This star has spectral type O9.5V and is a well-known runaway.}
{Spectral energy distributions of massive runaways are predicted for the whole electromagnetic spectrum.}
{
We conclude that the non-thermal radiation might be detectable at various energy bands for relatively nearby runaway stars, especially at high-energy gamma rays.  Inverse Compton scattering with photons from the heated dust gives the most important contribution to the high-energy spectrum. This emission approaches {{\it Fermi}} sensitivities in the case of $\zeta$ Oph.
}


\maketitle
%

\section{Introduction}

Runaway stars have high peculiar velocities, $v_{\star}$ $\sim$ 30 km s$^{-1}$ (e.g. Gies \& Bolton 1986; Tetzlaff, Neuh\"{a}user \& Hohle 2011). These stars move supersonically through the interstellar medium (ISM) and are ejected from their birth associations by gravitational kicks (e.g. Fujii \& Zwartz 2011). Runaway stars with strong winds -- spectral types OB and Wolf-Rayet (WR) -- can produce stellar bowshocks when interacting with the surrounding gas (e.g. Van Buren et al. 1995).  Bowshocks develop as arc-shaped structures, with bows pointing ahead of the stars (in the same direction as the stellar velocity).

There are several cataloged objects of this type (e.g. Van Buren et al. 1995;  Noriega-Crespo, van Buren, \& Dgani 1997;  Kobulnicky, Gilbert \& Kiminki 2009; Gvaramadze, Kniazev, \& Kroupa 2011; Peri et al. 2011).  Not all runaway stars produce a bowshock. Van Buren et al. (1995) concluded that only 30\% of the cataloged runaway stars present bowshock-like structures. More recently, Peri et al. (2011) found out that only $\sim$ 10\% of early-type runaway stars develop an observable bowshock in the infrared (IR). Some physical situations exist in which a bowshock does not form. If the star is moving in a low-density, hot plasma then it might be moving subsonically and a shock will not occur. Also, a bowshock might not form if the star has a weak wind or if  it has a high space velocity (see Comer\'on \& Kaper 1998, Huthoff \& Kaper 2002).

Bowshocks occur around many classes of astrophysical sources: pulsars, cataclysmic variables, colliding wind binaries, cometary H II regions, and even in groups and clusters of galaxies. A lot of  work has been done on bowshock modeling (e.g. Van Buren \& McCray 1988; Van Buren, Mac Low, Wood \& Churchwell 1990; Bandiera 1993; Van Buren 1993; Brighenti \& D\'ercole 1995; Wilkin 1996; Comer\'on 1997; Chen \& Huang 1997; Comer\'on \& Kaper 1998; Wilkin 2000; Wareing, Zijlstra \& O'Brien 2007). Related physical  problems such as  solving the astrophysical blunt-body problem are treated in e.g. Cant\'o \& Raga (1998) and more recently in  Schulreich \& Breitschwerdt (2011); the physics of mixing layers is discussed in e.g. Baranov, Krasnobaev \& Ruderman (1976), Cant\'o \& Raga (1991) and Raga, Cabrit \& Cant\'o (1995).

The supersonic stellar wind sweeps the ISM material and piles it up in the bowshock.
The stellar and shock-excited radiation heats this swept-up material. The dust, in turn, re-radiates the energy as mid-to-far IR flux (Van Buren \& McCray 1988).

Relativistic particles can be accelerated at strong shocks producing non-thermal emission (Drury 1983). Benaglia et al. (2010) have reported non-thermal radio emission from the bowshock of the runaway star BD +43$\degr$ 3654. This emission is thought to be synchrotron radiation generated by the interaction of relativistic electrons with the magnetic field of the source.

In this work we present a model for the radiative non-thermal emission that takes place in the  bowshocks of runaway stars. We compute the emission produced by the relativistic particles accelerated at the shock, and the thermal emission caused by free-free mechanism (Bremsstrahlung). Finally, we apply this model to the well-known stellar bowshock from the O9.5V star $\zeta$ Oph.
The bowshock shape is also computed, following the analytical method developed by Wilkin (2000).

Our paper is organized as follows. In the next section we present the radiative model developed to compute the non-thermal emission and the absorption that can take place in these types of sources. We also present a brief discussion on particle acceleration in shocks. In Sec. \ref{apli} we apply the model to a particular source, the bowshock from $\zeta$ Oph. We present the best-fit shell shape and the computed spectral energy distribution. Finally, in Sec. \ref{end} we discuss the detectability of the non-thermal radiation and offer our conclusions.

\section{Radiative model}\label{rad}

Bowshocks from runaway stars produce thermal emission by radiative heating of the swept-up dust by the stellar radiation field. The infrared signal of the heated dusty bowshock is strongest in the far infrared (e.g. Van Buren \& McCray 1988;  Kobulnicky, Gilbert \& Kiminki 2009). The dust temperature can be understood in terms of dust models. For a typical bowshock of a runaway star, a fraction $\sim$ $10^{-2}$ of the star bolometric luminosity is emitted in the infrared. In this paper we do not calculate this radiation, and leave it for a future work (del Valle et al., in preparation).

The shocked ISM can also produce thermal emission through free-free interactions (Bremsstrahlung). This emission  peaks at energies $\sim$ 1 eV; we calculate it in Sec.\ref{ther}. 

The interactions of locally accelerated relativistic particles with the matter, radiation and magnetic fields in the system produce non-thermal radiation. We calculate this  emission below. 

\subsection{Shocks and particle acceleration}\label{acc}

\begin{figure}
\begin{center}
\includegraphics[trim=0cm 0cm 0cm 0cm, clip=true, width=.8\hsize, angle=0]{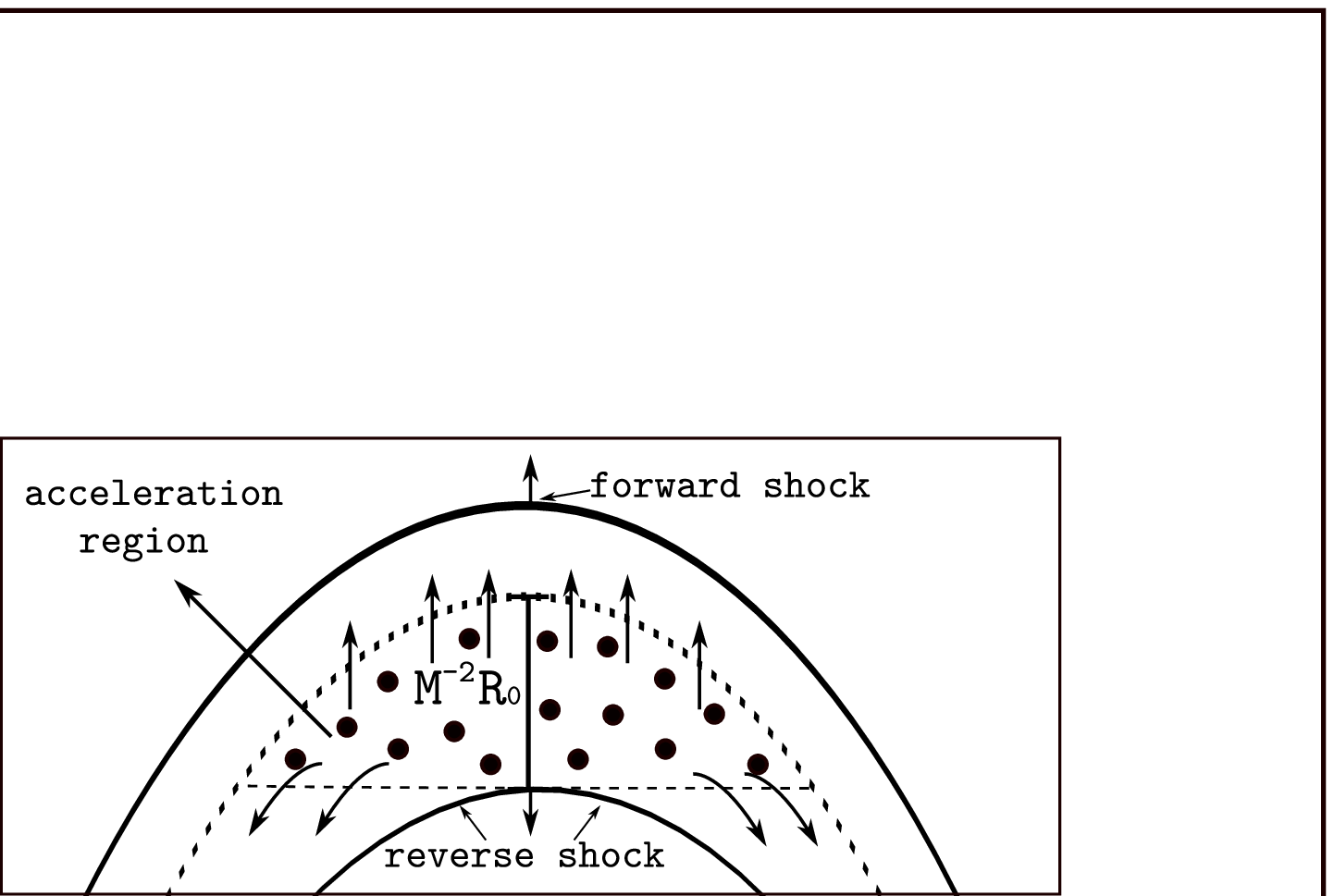}
\caption{Scheme of the system of shocks (not to scale).  The acceleration region is indicated by dots.}
\label{region}
\end{center}
\end{figure}
   
Shocks transfer  kinetic energy to non-thermal particles through particle acceleration. The acceleration mechanism is diffusive shock acceleration, the so-called Fermi I mechanism (e.g. Bell 1978). In this mechanism particles are accelerated by successive bouncing across the shock, gaining energy in each cross. The particle deflection is mediated by magnetic field irregularities. These irregularities are expected from turbulent  and magnetic instabilities.  For the mechanism to operate is necessary that the particles in the shocked medium (downstream) can effectively diffuse and reach the shock. 

The particle energy gain in each cycle (upstream-downstream-upstream) is ${\Delta}E/E$ $\propto$ $(v_{\rm s}/c)$, where $v_{\rm s}$ is the shock velocity; and after $k$ cycles the particle energy is $E = E_{\rm i}\left( 1+\frac{\Delta E}{E}\right)^{k}$, where $E_{\rm i}$ is the initial energy. The time that a  particle requires to accelerate up to an energy $E$ is given by
\begin{equation}
t_{\rm acc} = {\eta}\frac{E}{eBc}.
\label{acelera}
\end{equation}   
Here, $B$ is the magnetic field in the acceleration region, and $\eta$ is the acceleration efficiency (Drury 1983):
\begin{equation}
\eta \sim \frac{1}{10}\frac{r_{\rm g}c}{D}\left(\frac{v_{\rm s}}{c} \right)^{2}; 
\label{eta}
\end{equation}
$D$ is the diffusion coefficient, and $r_{\rm g} = E/(eB)$ is the particle gyroradius. In the Bohm limit $D_{B} = r_{\rm g}c/3$. Fast shocks are more efficient accelerators than slow shocks. The resulting spectrum of injected particles by this  mechanism is a power-law, i.e. $Q(E) \propto E^{-\alpha}$ (e.g. Protheroe 1999), with $\alpha \sim 2-2.2$. 

The collision of the supersonic stellar wind with the ISM around a runaway star results in a system of two shocks (see Fig. \ref{region}). In the steady state, mass and momentum are conserved  and a flow is established between the two shocks that carries away the mass and momentum deposited by the colliding fluids. In the reference frame of the star, the ISM can be considered as an incoming  wind of parallel streamlines and the stellar wind as a radial outflow (e.g. Wilkin 1996).
One of the shocks is a forward shock that propagates in the same direction as the stellar wind with velocity $v_{\rm fs}$ $\sim$ $V_\star$; and the other is a reverse shock that propagates in the opposite direction with $v_{\rm rs}$ $\sim$ $V_{\rm W}$. The stellar wind can be considered as a continuous power source, therefore both shocks reach a steady state. If radiative losses are inefficient within the shock discontinuity, the shock is adiabatic; otherwise, the shock is radiative. 

The stellar wind is divergent because its ram pressure decreases with distance, while the ram pressure of the ISM is constant. The point where the ram pressure of the wind and the ISM balance, i.e. $\rho_{\rm w}V_{\rm w}^{2} = \rho_{\rm a}V_{\star}^{2} $,  is called the standoff point, where 
${\rho}_{\rm w} = \dot{M}_{\rm w}/4{\pi}R^{2}V_{\rm w}$. This point of the shell defines  the standoff radius $R_{0}$:

\begin{equation}
R_{0} = \sqrt{ 
\frac{ \dot{M}_{\rm w} V_{\rm w}  }
{4 \pi {\rho}_{\rm a} V_{\star}^{2} }
}. 
\label{R0}
\end{equation}

The physical conditions in bowshocks from runaway stars produce an  adiabatic and fast reverse shock at the shocked stellar wind; and the forward  shock is radiative and slow (e.g. Van Buren 1993).  The radiative shock compresses the material, and as the temperature decreases, the density grows.

Kis et al. (2004) demonstrated that upstream particles in the Earth bowshock undergo diffusive transport into the upstream region, a direct evidence of Fermi I acceleration (e.g. Burgess 2007). Benaglia et al. (2010) reported the detection of non-thermal radio emission from the bowshock of the runaway star BD+43$\degr$3654. The non-thermal radiation is expected from synchrotron emission generated by relativistic electrons accelerated  at either the forward shock or the reverse shock.   

We considered that an initial supra-thermal population of relativistic particles, electrons and protons, are accelerated in the adiabatic shock (reverse shock). This shock is faster than the forward shock, so in the reverse shock the relativistic particles are accelerated more efficiently by the Fermi mechanism (see Eq. (\ref{eta})). All  equations given above are valid for planar shocks. We considered an acceleration region where the shock is nearly flat (see Fig.\ref{region}). The width of the shocked stellar wind, $\Delta$, can be estimated as $\Delta$ $\sim$ $M^{-2}R_{0}$, where $M$ is the Mach number of the shocked wind.  The acceleration region is assumed as a small region near the bowshock apex, of scale length $\sim$ $\Delta$; the acceleration volume is given in Table \ref{paraRad}.

The available power in the system is the kinetic power from the stellar wind: 
\begin{equation}
L_{\rm T} \sim  \frac{1}{2}{\dot M_{\rm w}}V_{\rm w}^2. 
\end{equation}

To estimate the magnetic field in the flow, we consider  that the magnetic energy density is in  subequipartition with respect to the kinetic energy $L_{\rm T}$,  by a 0.1 factor, i.e.:
\begin{equation}
\frac{B^{2}}{8\pi} = \frac{0.1 L_{\rm T}}{V_{\rm w} A},
\label{equipar}
\end{equation}
where $A$ is the area of a sphere of radius $R_{0}$. This condition ensures that the flow is matter-dominated, i.e. compressible, allowing shocks to develop.

The kinetic power available in the acceleration region is $L = f L_{\rm T}$, where $f$ is the ratio of the volume of a sphere of radius $R_{0}$ and the volume of the acceleration region. A small fraction of this kinetic power goes into relativistic particles. $L_{\rm rel} = q_{\rm rel} L$. We adopted a standard fraction $q_{\rm rel} = 0.1$ (e.g. Protheroe 1999). We took into account both hadronic and leptonic content in the relativistic power, $L_{\rm rel} = L_{\rm p} + aL_{\rm e}$. The ratio of relativistic protons to electrons, ${\it a}$,  is unknown. We considered two cases $ a = 1$ (equal energy density in both species)  and $a = 100$ (as observed in Galactic cosmic rays, Ginzburg \& Syrovatskii 1964). 

Here we discuss two different types of massive star: an O4I and an O9I star. The values for the parameters  involved are given in Table \ref{paraRad}.

\begin{table*}
\begin{center}

\begin{tabular}{lll}
\hline\noalign{\smallskip}
 Parameter & O4 & O9 \\[0.001cm]
\hline\noalign{\smallskip}
$R_{0}$  Standoff radius [pc] & 8.3 & 0.2\\[0.001cm]
$V_{\star}$  Spatial velocity [km s$^{-1}$] & 100 &  30\\[0.001cm]
$V_{\rm w}^{\rm a}$  Wind velocity [km s$^{-1}$] & 2.2$\times 10^3$ & 0.8$\times 10^3$   \\[0.001cm]
${\dot{M}}_{\rm w}^{\rm a}$  Wind mass loss rate [M$_{\odot}$ yr$^{-1}$] & $10^{-4}$ & $10^{-6}$ \\[0.001cm]
$n_{\rm a}$ Ambient density [cm$^{-3}$] & 1 & 100 \\[0.001cm]
$B^{\rm b}$ Magnetic field [G] & $\sim$ 3.0$\times$10$^{-5}$ & $\sim$ 10$^{-5}$\\[0.001cm]
$\eta$  Acceleration efficiency & $\sim$ 2.0$\times 10^{-5}$ & $\sim$ 2.7$\times 10^{-6}$\\[0.001cm]
$L$   Available power [erg s$^{-1}$] & $\sim$ 3.2$\times10^{36}$  & $\sim$ 4.3$\times 10^{33}$\\[0.001cm]
$a$ Hadron-to-lepton energy ratio & 1 & 100\\[0.001cm]
$q_{\rm rel}$  Jet content of relativistic particles & 10\% & 10\% \\[0.001cm]
$\alpha$  Injection index & 2 & 2 \\[0.001cm]
${\Delta}$  Thickness of shocked wind [$R_{0}$] & $\sim$ 0.3 & $\sim$ 0.3  \\[0.001cm]
Vol$_{\rm acc}$  Acceleration region volume [cm$^{-3}$] & $\sim$ 7$\times 10^{56}$  & $\sim$ $10^{51}$ \\[0.001cm]
${L_{\star}}^{\rm c}$ Star luminosity [$L_{\odot}$] & $\sim$ 7$\times 10^{5}$ & $\sim$ 5$\times 10^{4}$  \\[0.001cm]
 ${T_{\rm IR}}$ Dust temperature [K] & $\sim$ 24 &  $\sim$ 54  \\[0.001cm]
\hline\\[0.05cm]
\end{tabular}	
\caption[]{Parameters for the different types of stars.\\
$^{\rm a}$Values derived by  Kobulnicky, Gilbert \& Kiminki (2009).\\
$^{\rm b}$This value corresponds to the magnetic field in the acceleration region, obtained from Eq. (\ref{equipar}).\\
$^{\rm c}$Values from Martins, Schaerer \& Hillier (2005).
}
\label{paraRad}
\end{center}
\end{table*}

\subsection{Non-thermal radiative losses}

The electrons lose energy mainly by inverse Compton (IC) scattering, synchrotron radiation, and relativistic Bremsstrahlung.

The synchrotron cooling time  rate is
\begin{equation}
t_{\rm sy}^{-1} = \frac{4}{3}\frac{\sigma_{\rm T}cU_{B}}{m_{\rm e}c^{2}}\biggl(\frac{m_{\rm e}}{m}\biggr)^{3}\frac{E}{mc^{2}},
\end{equation}
where $\sigma_{\rm T}$ is the Thomson cross section and $U_{B}$ is the magnetic energy density.

The IC losses can be calculated from (Blumenthal \& Gould 1970)
\begin{equation}
t_{\rm IC}^{-1} = \frac{1}{E_{\rm e}}\int_{\epsilon_{\rm min}}^{\epsilon_{\rm max}}
\int_{\epsilon}^{\frac{bE_{\rm e}}{1+b}} (\epsilon_{1} - \epsilon) \frac{{\rm d}N}{{\rm d}t{\rm d}\epsilon_{1}}{\rm d}{\epsilon}_{1} , 
\end{equation}
where $\epsilon$ and  $\epsilon_{1}$ are the incident photon and scattered photon energy, respectively, and
 
\begin{equation}
\frac{{\rm d}N}{{\rm d}t{\rm d}\epsilon_{1}} = \frac{1}{E_{\rm e}}\frac{2\pi r_{0}^{2}mc^{3}}{\gamma}\frac{n_{\rm ph}(\epsilon){\rm d}\epsilon}{\epsilon}f(q)
\end{equation}
with
\begin{equation}
f(q) = 2q\ln{q}+(1+2q)(1-q)+\frac{1}{2}\frac{(bq)^{2}}{a+bq}(1-q),
\label{efe}
\end{equation}
where $b = 4\epsilon\gamma/mc^{2}$  and $q = \epsilon_{1}/[b(E_{\rm e}-\epsilon_{1})]$. This expression takes into account losses in the Klein-Nishina regimen. The target photon fields  are the radiation fields of the acceleration region. The fields considered here are the stellar radiation field at a distance $R_{0}$ from the star -- assumed as a black body at temperature $T_{\star}$-- , and the IR radiation from the heated dust, also considered as a black body at  $T_{\rm IR}$. Dust grains are heated by starlight, and cool by radiating in the infrared. To estimate $T_{\rm IR}$ we used a simplified dust model based on Draine \& Lee (1984): 
\begin{equation}
T_{\rm IR} = 27\,a_{\mu {\rm m}}^{-1/6}\,L_{{\star} 38}^{1/6}\,R_{0 {\rm pc}}^{-1/3} \,\, {\rm K}
\label{Temp}
\end{equation}
(e.g. Van Buren \& McCray 1988). Here $a_{\mu {\rm m}}$ $\sim$ $0.2$ $\mu$m is the dust grain radius. More detailed and complex dust emission models are beyond the scope of this work. For more details see e.g. Draine \& Li (2007) and Draine (2011).

The relativistic Bremsstrahlung losses are calculated considering a complete ionized plasma, using (Berezinskii et al. 1990)
\begin{equation}
t_{\rm Br}^{-1} = 4nZ^{2}r_{e}^{2}\alpha c \biggl[\ln{\frac{2E_{\rm e}}{m_{\rm e}c^{2}}}-\frac{1}{3}\biggr],
\end{equation}
where $n$ is the density of target ions in the acceleration region (the shocked stellar wind in this case). The density of the shocked wind  according to the Rankine-Hugoniot equations for adiabatic shocks (e.g. Landau \& Lifshitz 1959) is 4$n_{\rm w}$, where $n_{\rm w}$ is the wind density.
 
Protons  lose energy through proton-proton inelastic collisions with the shocked wind material. These interactions produce neutral and charged pions; the former decay and  produced gamma rays (e.g. Vila \& Aharonian 2010). The latter decay into secondary electrons/positrons and neutrinos. The losses produced by proton-proton interactions occur on time scales of
\begin{equation}
t_{pp}^{-1} = n c\sigma_{pp}K_{pp}, 
\end{equation}
where $n$ is the density of target protons, $\sim 4 n_{\rm w}$, and $K_{pp}$ is the ineslasticity ($\sim$ 0.5). The cross section can be approximated by (Kelner, Aharonian, \& Bugayov 2006)
\begin{equation}
\sigma_{pp} = (34.3+1.88L+0.25L^{2})\biggl[1-\biggl(\frac{E_{\rm th}}{E_{p}}\biggr)^{4}\biggr]^{2},
\end{equation}
where $L = \ln(E_{p}/1{\rm TeV})$; and $E_{\rm th} = 280$ MeV.

\begin{figure*}
\begin{centering}
\resizebox{.45\hsize}{!}{\includegraphics[trim=0cm 0cm 0cm 0cm, clip=true,angle=270]{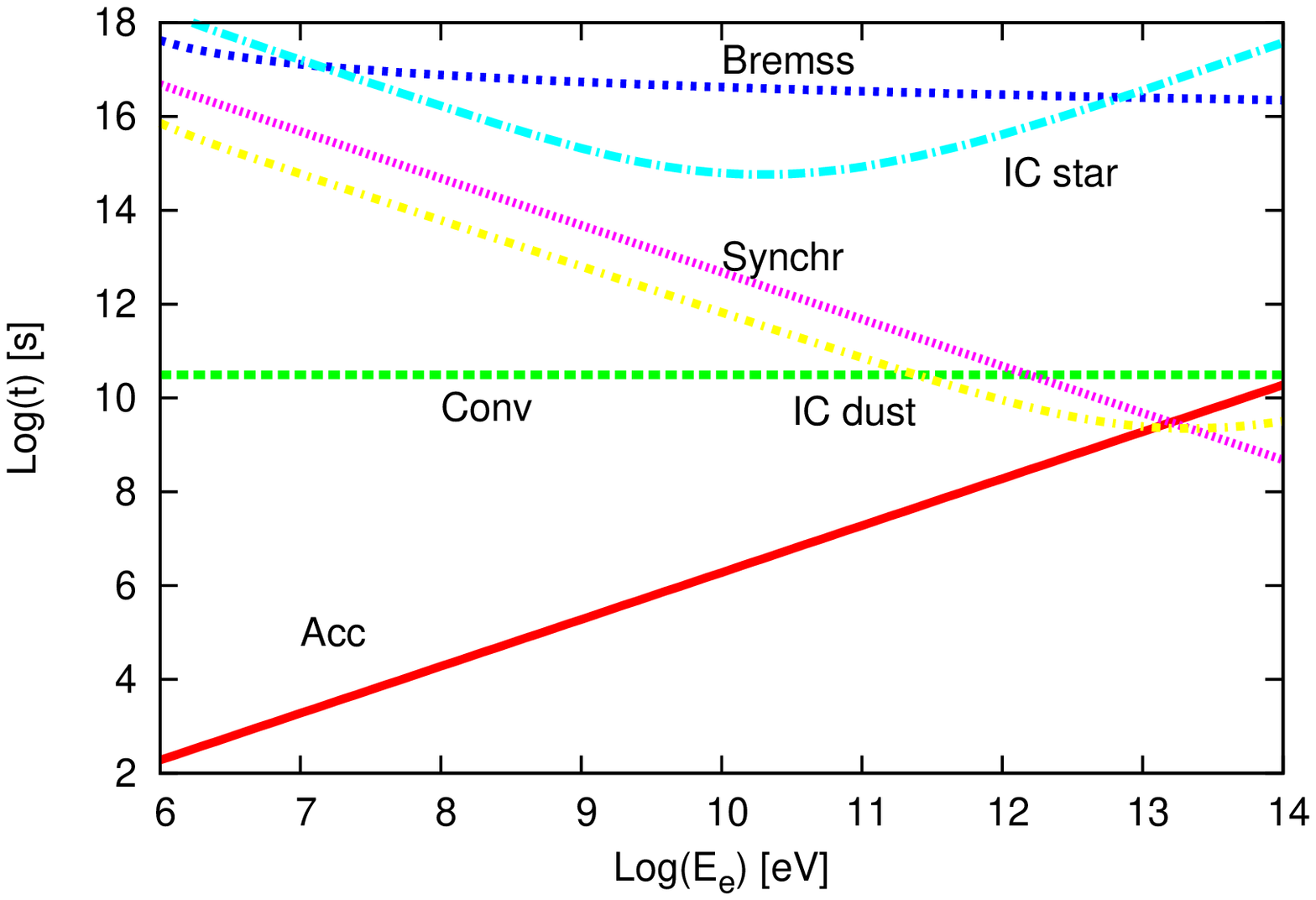}}
\resizebox{.45\hsize}{!}{\includegraphics[trim=0cm 0cm 0cm 0cm, clip=true,angle=270]{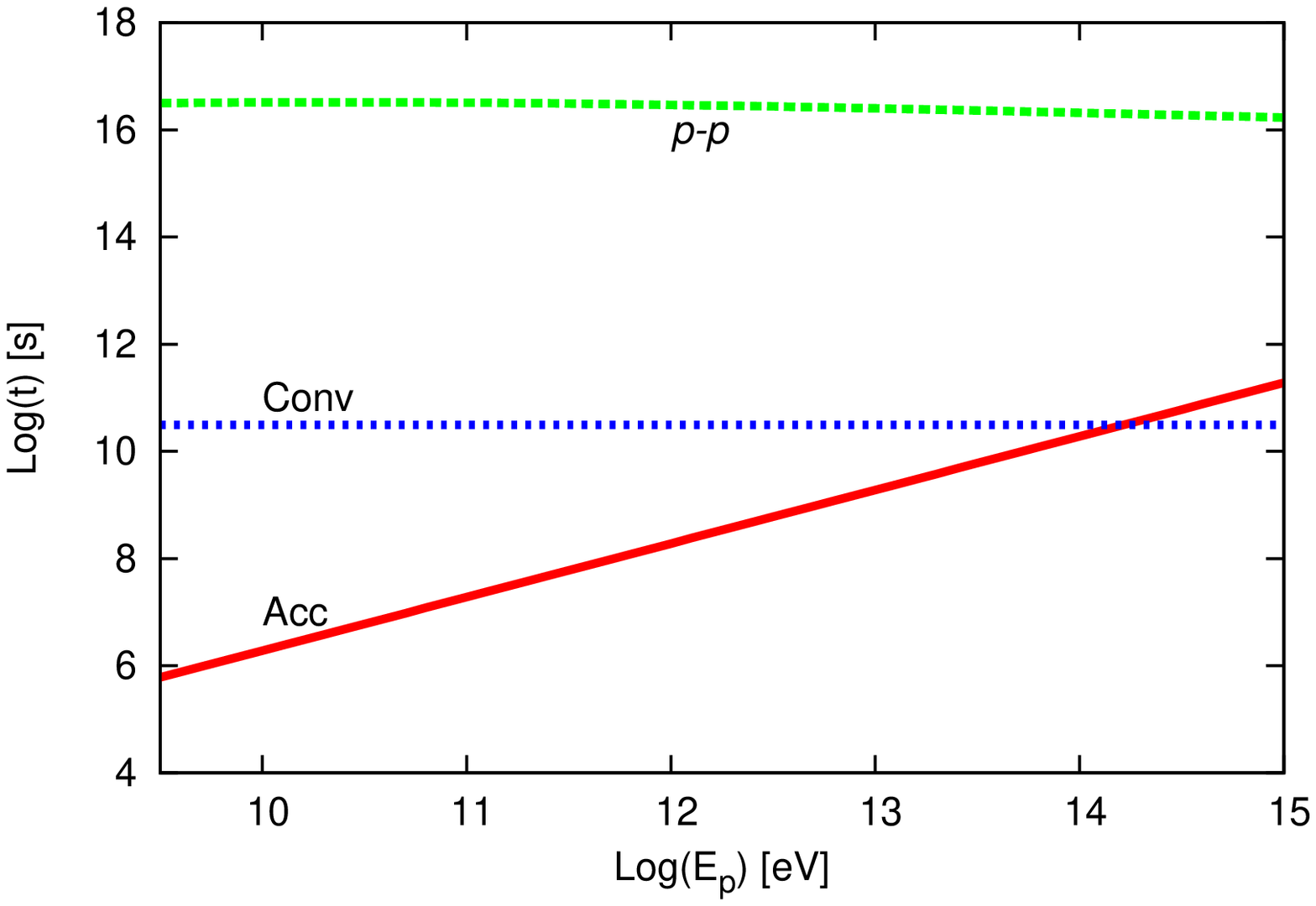}}
\resizebox{.45\hsize}{!}{\includegraphics[trim=0cm 0cm 0cm 0cm, clip=true,angle=270]{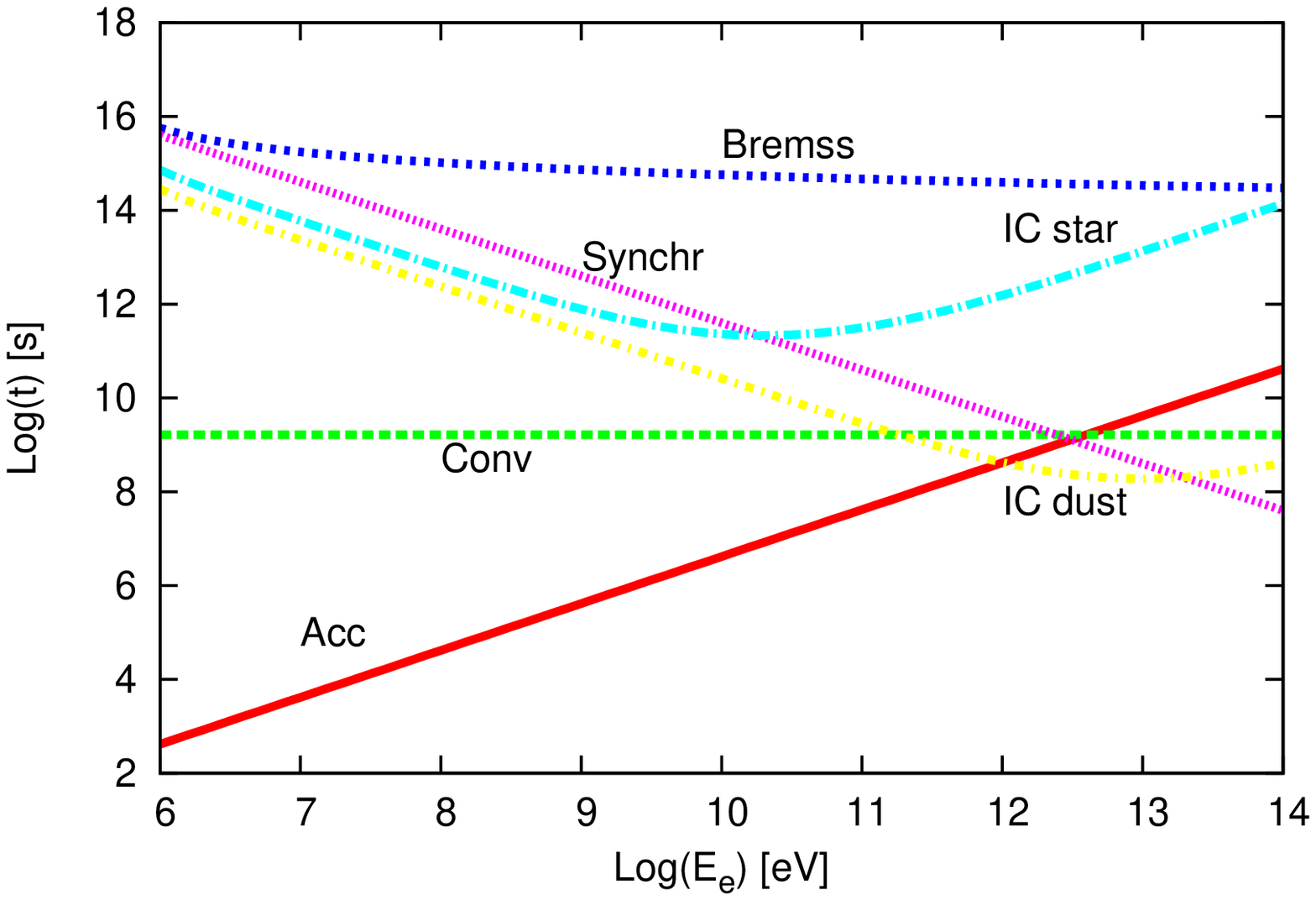}}
\resizebox{.45\hsize}{!}{\includegraphics[trim=0cm 0cm 0cm 0cm, clip=true,angle=270]{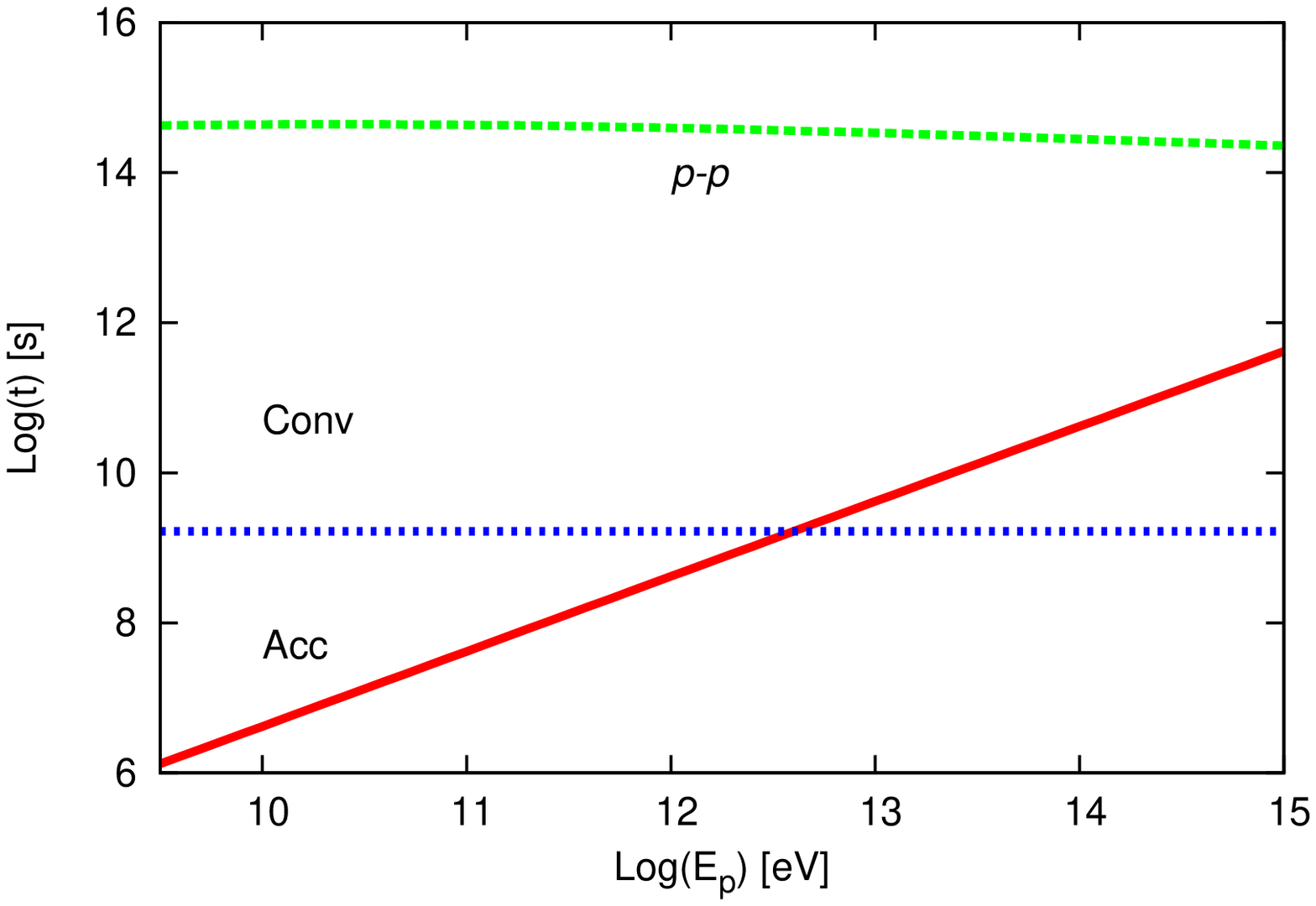}}
\caption{Acceleration and cooling time scales for electrons and protons for an O4I star (up), and for an O9I star (down). Left panels are for electrons and right panels are for protons. The non-radiative losess produced by convection are also shown.}
\end{centering}
\label{fig:cool}
\end{figure*}
Photomeson production is irrelevant at the energies considered in this paper.

In Fig. \ref{fig:cool} we show the cooling rates for electrons and protons in the acceleration region for an O4I  and an O9I star. In the O4I star system the IC scattering of IR photons dominates the radiative losses. For the O9I star, the IC scattering of IR photons and synchrotron radiation prevail among the radiative losses.  

The relativistic particles can also suffer from non-radiative losses due to escape from the acceleration region. Particles can be convected away by the stellar wind on a time $t_{\rm conv}$ $\sim$  $\Delta/V_{\rm w}$. This non-radiative loss dominates the proton cooling rate in both types of runaway stars (O4I and O9I).

The minimum kinetic energy for each particle is considered to be on the order of the rest mass energy. The maximum energy for the electrons and protons  is obtained equating the smallest cooling rate to the acceleration rate, given by Eq. (\ref{acelera}).  In the O4I system  the electrons reach energies  $\sim$ 10 TeV  and the protons can reach  energies $\sim$ 10$^{2}$ TeV. For O9I electrons  reach energies $\sim$ TeV, and  protons energies $\sim$ 10 TeV.  In both cases the Hillas criterion is satisfied:

\begin{equation}
E_{\rm max} <  300\,(r_{\rm g}/{\rm cm}) \, (B/{\rm G})\,{\rm eV}, 
\end{equation}
where $r_{\rm g}$ is the maximum particle gyroradius in the available space, i.e. $\sim$ ${\Delta}$.   

\subsection{Particle distributions}

To calculate the steady state particle distributions $N(E)$ for electrons and protons, we solved the transport equation in steady state (Ginzburg \& Syrovatskii 1964):
\begin{equation}
 \frac{\partial}{\partial E}\biggl[\frac{{\rm d}E}{{\rm d}t}{\bigg\arrowvert}_{\rm loss}N(E)\biggr]+\frac{N(E)}{t_{\rm esc}} = Q(E),
\label{tra}
\end{equation}
where $t_{\rm esc}$ is the convection time. We assumed that all physical properties in the acceleration region are homogeneous.

$Q(E)$ is the injection function, a power-law in the energy of the particles, as expected from  diffusive shock acceleration:
\begin{equation}
Q(E)=Q_{0}\,E^{-\alpha}.
\end{equation}

The normalization constant $Q_{0}$ for each type of particles is obtained from $L_{{\rm e},p}$ as
\begin{equation}
L_{{\rm e},p}  = V \int_{E_{{\rm e},p}^{\rm min}}^{E_{{\rm e},p}^{\rm max}}
{\rm d}E_{{\rm e},p}E_{{\rm e},p} Q_ {{\rm e},p}(E_{{\rm e},p}).
\end{equation}
Here $V$ is the volume of the acceleration region. 

The exact solution to the equation is also a power-law: 
\begin{eqnarray}
N(E)= &&\biggl\arrowvert \frac{{\rm d}E}{{\rm d}t}\biggl{\arrowvert}_{\rm loss}^{-1}\int_{E}^{E^{\rm max}}
{\rm d}E' Q(E')\nonumber\\ &&\times{\rm exp}\biggl(-\frac{\tau(E,E')}{t_{\rm esc}}\biggr),
\end{eqnarray}
with
\begin{equation}
\tau(E,E')= \int_{E}^{E'} {\rm d}E'' \biggl\arrowvert \frac{{\rm d}E''}{{\rm d}t}\biggr{\arrowvert}_{\rm loss}^{-1}. 
\end{equation}

\subsection{Non-thermal radiative processes}

As mentioned before, the interactions of the relativistic particles with the various fields in the system produce non-thermal emission. 

We considered synchrotron emission, inverse Compton scattering with the stellar photon field and the IR emission, and relativistic Bremsstrahlung from electrons. For protons,  inelastic collisions between the relativistic protons with  the shocked wind material were also calculated. 

The synchrotron emission is given by
\begin{eqnarray}
L_ {\gamma}(E_ {\gamma}) = && {\kappa}_{\rm SSA}(E_{\gamma})E_ {\gamma} V \frac{\sqrt{3}e^{3}B}{hmc^{2}}\int_{E_{\rm min}}^{E_ {\rm max}} 
{\rm d}E N(E)\frac{E_ {\gamma}}{E_ {\rm c}}\nonumber\\&&1.85\times\biggl(\frac{E_ {\gamma}}{E_ {\rm c}}\biggr)^{1/3}{\exp}\biggl(\frac{E_ {\gamma}}{E_ {\rm c}}\biggr),
\label{Syn}
\end{eqnarray}
where
\begin{equation}
E_ {\rm c} = \frac{3}{4\pi}\frac{ehB}{mc}\biggl(\frac{E}{mc^{2}}\biggr)^{2}.
\end{equation}
$\kappa_{\rm SSA}$ is the synchrotron self-absorption (SSA) factor:
\begin{equation}
\kappa_{\rm SSA} = \frac{1-e^{-\tau_{\rm SSA}(E_{\gamma})}}{\tau_{\rm SSA}(E_{\gamma})};
\end{equation}
here $\tau_{\rm SSA}(E_{\gamma})$ is the SSA optical depth (see Rybicki \& Lightman 1979).

The IC emission is calculated using
\begin{eqnarray}
L_{\gamma}(E_ {\gamma}) = && E_{\gamma}^{2} V \int_{E_{\rm min}}^{E_{\rm max}}{\rm d}E_{\rm e}N_{\rm e}(E_{\rm e})\nonumber\\&&\times\int_{\epsilon_{\rm min}}^{\epsilon_{\rm max}}{\rm d}\epsilon
P_ {\rm IC}(E_{\rm e},E_{\gamma},\epsilon).
\label{IC}
\end{eqnarray}
The spectrum of scattered photons is 
\begin{equation}
P_{\rm IC}(E_{\rm e},E_{\gamma},\epsilon) = \frac{3\sigma_{\rm T}c(m_{\rm e}c^{2})^{2}}{4E_{\rm e}^{2}}
\frac{n_{\rm ph}(\epsilon)}{\epsilon} f(q),
\end{equation}
with $f(q)$ given by Eq. (\ref{efe}).

The relativistic Bremsstrahlung luminosity is
\begin{equation}
L_{\gamma}(E_{\gamma})= E_{\gamma}V \int_{E_{\gamma}}^{\infty} n \sigma_{\rm B}(E_{\rm e},E_{\gamma})
\frac{c}{4\pi}N_{\rm e}(E_{\rm e}){\rm d}E_{\rm e},
\end{equation}
where 
\begin{equation}
\sigma_{\rm B}(E_{\rm e}, E_{\gamma}) = \frac{4\alpha r_{\rm 0}^{2}}{E_{\gamma}}\phi(E_{\rm e}, E_{\gamma}),
\end{equation}
and 
\begin{eqnarray}
\phi(E_{\rm e}, E_{\gamma}) = &&[1+(1-E_{\gamma}/E_{\rm e})^{2}-2/3(1-E_{\gamma}/E_{\rm e})]\nonumber\\ 
&&\times\biggl\{\ln{\frac{2E_{\rm e}(E_{\rm e}-E_{\gamma})}{m_{\rm e}c^{2}E_{\gamma}}} -\frac{1}{2}\biggr\}.
\end{eqnarray}

To  compute the gamma-ray emission produced by neutral pion decays  for $E_{p} <$ 0.1 TeV, we used the following expression:
\begin{equation}
L_{\gamma}(E_{\gamma}) = 2 V E_{\gamma}^{2}  \int_{E_{\min}}^{\infty} \frac{q_{\pi}(E_{\pi})}{\sqrt{E_{\pi}^{2}-m_{\pi}^{2}c^{4}}} {\rm d} E_{\pi},
\end{equation}
with $E_{\rm min}= E_{\gamma}+m_{\pi}c^{4}/4E_{\pi}$.
The $\pi^{0}$-emissivity is given by (Aharonian \& Atoyan 2000)
\begin{equation}
q_{\pi}(E_{\pi}) = \frac{n_{\rm p}}{\kappa_{\pi}}\sigma_{pp} \biggl(m_{p}c^{2}+E_{\pi}/\kappa_{\pi} \biggr)J_{p}\biggl(m_{ p}c^{2}+E_{\pi}/\kappa_{\pi}\biggr)
\end{equation}
with $\kappa_{\pi} \sim $ 0.17 (Gaisser 1990). For $E_{ p} <$ 0.1 TeV down to the threshold energy, the following replacement is necessary
\begin{equation}
\delta(E_{\pi}-{\kappa}_{\pi}E_{\rm kin}) \rightarrow \tilde{n}\delta(E_{\pi}-{\kappa}_{\pi}E_{\rm kin}).
\end{equation}
Here $\tilde{n}$ is the total number of $\pi^{0}$ created per $p-p$ collision, and $E_{\rm kin} = E_{p} - m_{p}c^{2}$ is the proton kinetic energy.

The gamma-ray luminosity in the range 0.1 TeV$\le E_{p} \le 10^{5}$ TeV can be obtained from (Kelner et al. 2006):
\begin{eqnarray}
L_{\gamma}(E_{\gamma}) = && n E_{\gamma}^{2} V \int_{E_{\gamma}}^{\infty} \sigma_{\rm inel}(E_{p})
N_{p}(E_{p})\nonumber\\&&\times F_{\gamma}\biggl(\frac{E_{\gamma}}{E_{p}},E_{p}\biggr) \frac{{\rm d}E_{p}}{E_{p}},
\end{eqnarray}
with $F_{\gamma}\biggl(\frac{E_{\gamma}}{E_{p}},E_{p}\biggr)$ a function of $E_{\gamma}$ and $E_{p}$ (see Vila \&  Aharonian 2009 and references therein).  

Secondary electron-positron pairs are created through the $p-p$ inelastic collisions (see Orellana et al. 2007). These particles can also produce non-thermal emission, which we took into account. For the relevant formulae see Orellana et al.'s paper.

\begin{figure*}
\resizebox{.45\hsize}{!}{\includegraphics[angle=270]{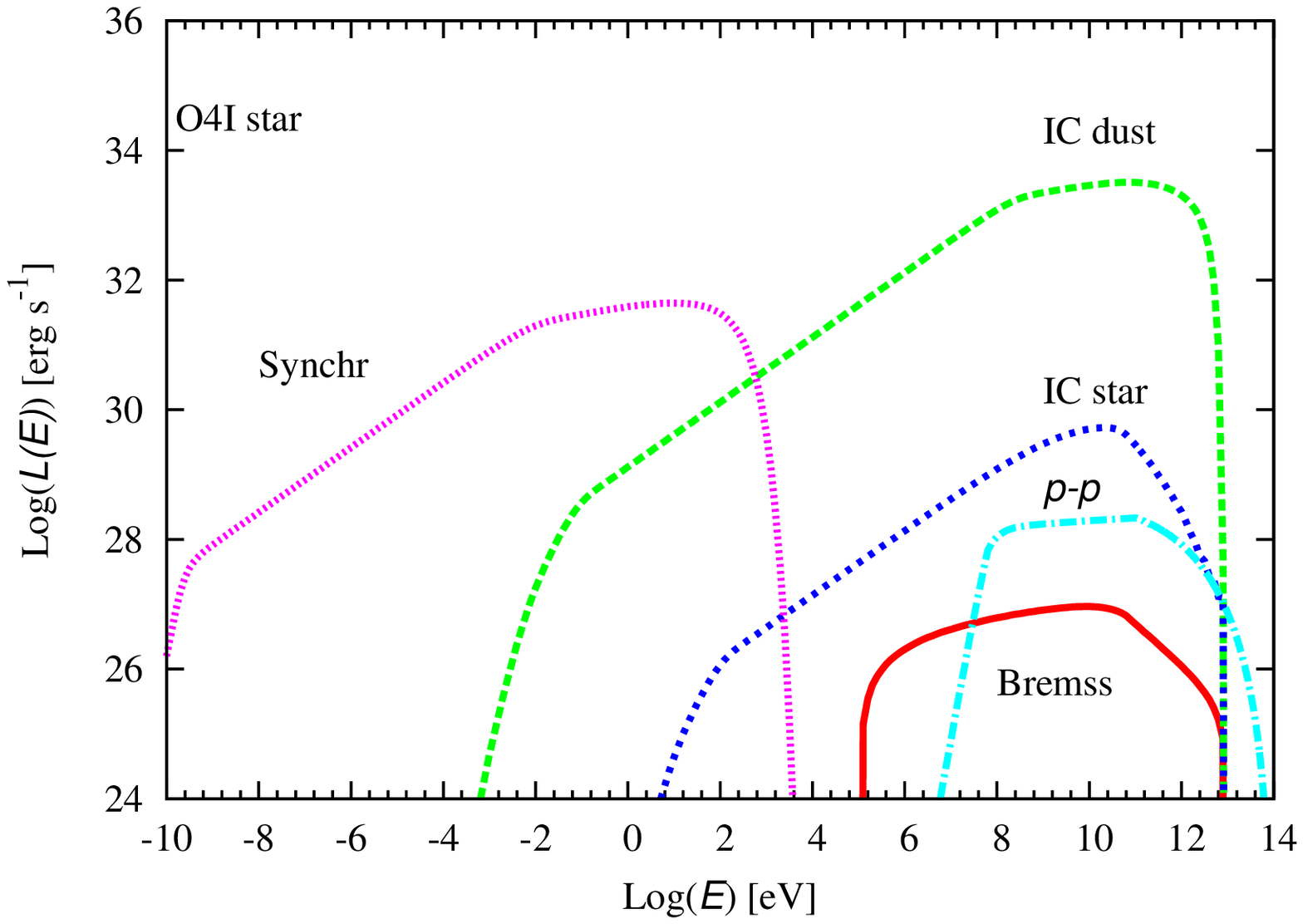}}
\resizebox{.45\hsize}{!}{\includegraphics[angle=270]{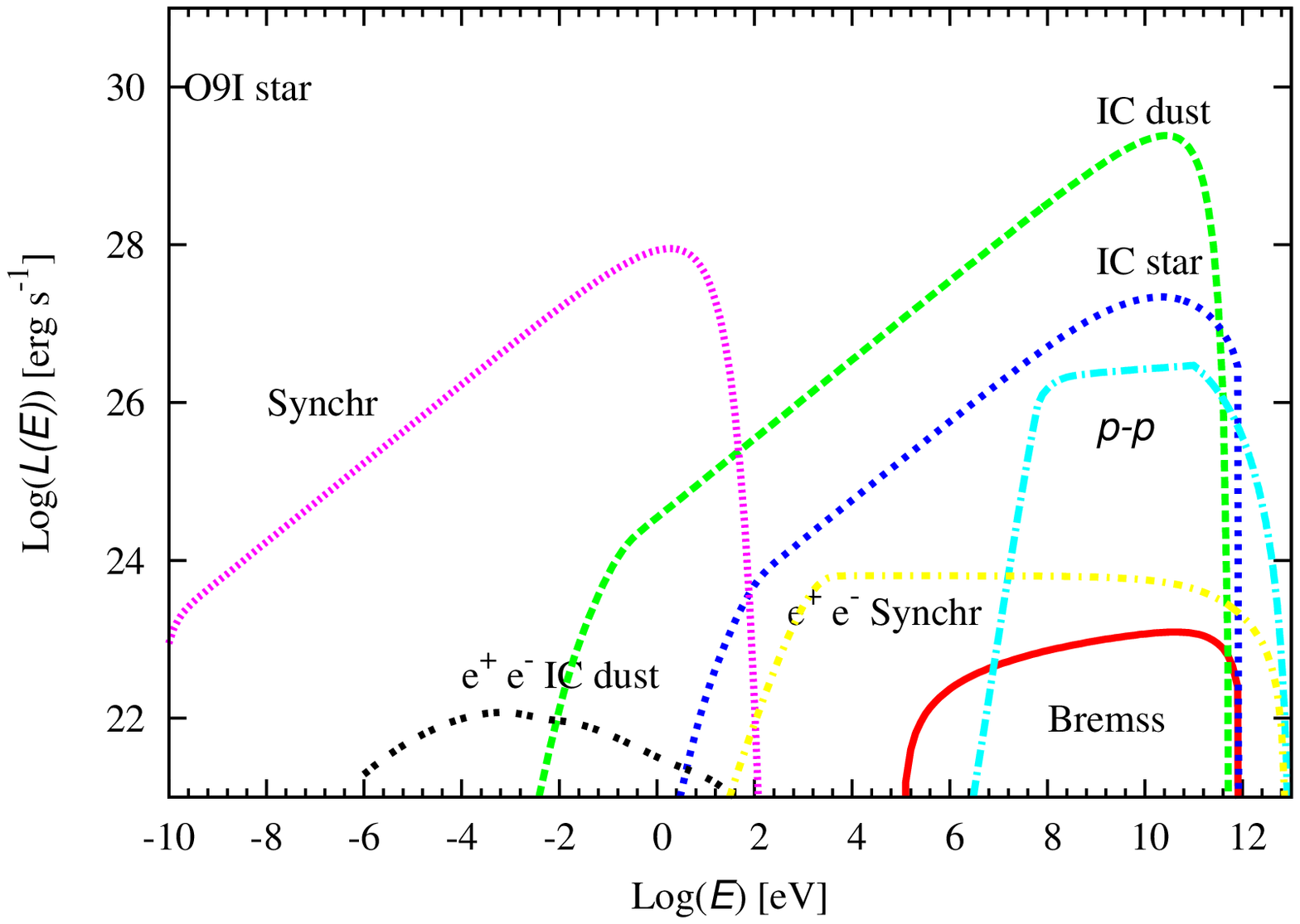}}
\caption{Non-thermal emission from the shocked stellar wind for an O4I star (left), and for an O9I star (right).}
\label{SEDs}
\end{figure*}

Figure \ref{SEDs} shows the computed SED for both types of stars. We calculated the contribution of the secondary pairs, to synchrotron emission and IC luminosity, only for the O9I star (because we assumed a proton dominated content in this case). The non-thermal luminosity of the O4I star is higher, mainly because the available power in the system is greater. The IC emission produced in the interaction with the dust photons dominates both SEDs at high energies. At lower energies synchrotron radiation dominates, but is relatively more important in the O9I system.

The IC emission from the dust photons, in general, increases as the temperature of the dust also  increases. This temperature depends, as a first aproximation, on $L_{\star}$ and $R_{0}$ (see Eq. (\ref{Temp})). As argued in Sec. \ref{acc},  $B$ must be at subequipartition values. This parameter not only affects the synchrotron emission, but the acceleration rate (see Eq. (\ref{acelera})). A higher value of $B$ gives a lower acceleration rate and the maximum particle energies change (decrease).

\subsection{Emission of the shocked ISM}\label{diff}

\subsubsection{Non-thermal emission}

Some relativistic particles can diffuse from the acceleration region to the shocked ISM before losing most of their energy. The diffusion time can be estimated as ${\Delta}^2/2D_{\rm B}$, where $D_{\rm B}$ is the diffusion coefficient in the Bohm limit as before. Particles that  diffuse to the denser shocked ISM (see Fig. \ref{region}) can interact there through $p-p$ inelastic collisions and relativistic Bremsstrahlung. The shocked ISM is contained within a very thin layer.

Because of the radiative losses the temperature  decreases and the density increases in the shocked ISM; both quantities are described by a temperature profile $T_{\rm fs}(x)$ and a density profile $n_{\rm fs}(x)$. 
The temperature profile is  given by (Zhekov \& Palla 2007)

\begin{equation}
T_{\rm fs} = \left( -7\times 10^{-19}\frac{2 P_{\rm ad}}{5 n_{\rm a}v_{\star}k_{\rm B}^3}(3.6)x + T_{\rm ad}^{3.6}\right)^{1/(3.6)},  
\label{Tpro}
\end{equation}
where $x$ is the distance from the shock; $T_{\rm ad} \sim 2 \times 10^9 v_{\star}^2$ K and $P_{\rm ad} = \frac{3}{4} {\rho_{\rm a}}v_{\star}^2$, are the corresponding   values of temperature and pressure for the adiabatic case. 

The density profile is
\begin{equation}
n_{\rm fs}(x) = \frac{C_{0} P_{\rm ad}}{k_{\rm B}T_{\rm fs}(x)},
\label{npro}
\end{equation}
with $C_{0}$ a normalization constant such that $n_{\rm a} \equiv n_{\rm fs}(T_{\rm a})$, where  $T_{\rm a}$ is the ambient temperature. We estimated the width of the shocked region as $x_{\rm c}$, where $T(x_{\rm c})= T_{\rm a}$.

\begin{figure}
\begin{centering}
\resizebox{.7\hsize}{!}{\includegraphics[angle=270]{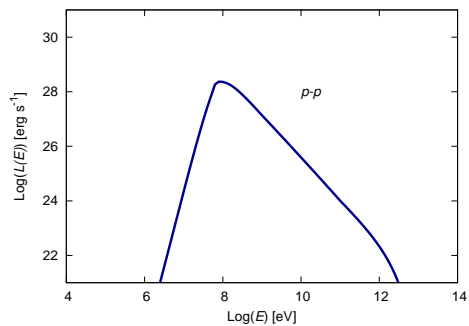}}
\caption{${\gamma}-$ ray emission from the shocked ISM produced by $p-p$ interactions from an O9I star.}
\label{Dif}
\end{centering}
\end{figure}

The particle energy distribution of the diffused protons is obtained solving Eq. (\ref{tra}), considering a 3-D delta function injection (see Aharonian \& Atoyan 1996; Bosch-Ramon, Aharonian \& Paredes 2004):
\begin{equation}
 Q(E,R,t) = N_{\rm p} \delta({\bf R}) \delta(t),
\end{equation}
where $N_{\rm p}$ is the initial energy distribution of the particles accelerated at the reverse shock.
The general solution is 
\begin{eqnarray}
N_{\rm p \,diff}(E)= && \frac{KE^{-\beta}\exp{-E/E_{\rm max}}}{\pi_{3/2}R_{\rm diff}^{3}}\nonumber\\ &&\times{\rm exp}\biggl(-\frac{(\beta-1)t}{\tau} - \frac{R^2}{R_{\rm diff}^{2}}\biggr),
\end{eqnarray}
where $\tau_{\rm pp}$ $\sim$ $6 \times 10^{7} (\bar{n}/1 \, {\rm cm}^{-3})^{-1}$ yr. 
Here we consider an average value for the shocked density $\bar{n}_{\rm fs}$ $=$ $\int_{0}^{x_{\rm c}} n_{\rm fs}(x) dx/x_{\rm c}$; $t$ is the propagation time after the injection into the shocked ISM, $t$ $\sim$ $c x_{\rm c}$. $R_{\rm diff}$ is the so-called diffusion radius. For $t$ $<<$ $\tau_{pp}$,  $R_{\rm diff} = 2 \sqrt{Dt}$, with $D$ the diffusion coefficient, we considered $D = D_{\rm B}$, see Sec. \ref{acc}. 

The particle energy distributions of the diffused electrons can be approximated by
\begin{equation}
N_{\rm e \, diff} \sim Q(E,R,t) \; t_{\rm cool},
\end{equation}
where $1/t_{\rm cool} = 1/t_{\rm synchr}^{-1}+t_{\rm IC}^{-1} + t_{\rm Bremss}^{-1}$. 
The interactions of diffused particles with the matter of the shocked ISM -through $p-p$ interactions for protons and relativistic Bremsstrahlung for electrons- produce non-thermal emission. 

For the two stars considered here, only the protons from the O9I system can diffuse into the  shocked ISM. Fig. \ref{Dif} shows this contribution to the non-thermal SED. This emission peaks at $E$ $\sim$ 100 MeV, with $L$ $\sim$ $10^{28}$ erg s$^{-1}$.  From Fig. \ref{SEDs} it can be seen that this contribution is negligible compared to the total IC radiation.     

\subsubsection{Thermal emission}\label{ther}

\begin{figure}
\begin{centering}
\resizebox{.7\hsize}{!}{\includegraphics[trim=0cm 0cm 0cm 0cm, clip=true,angle=270]{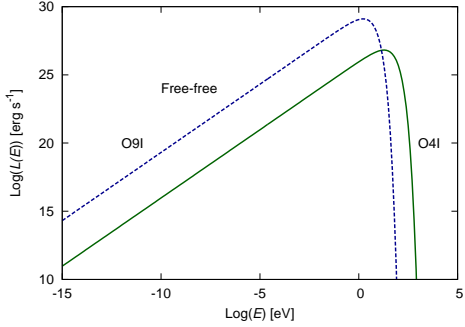}}
\caption{Thermal emission from the shocked ISM produced by free-free mechanism for an O4I star and for an O9I star.}
\label{Therm}
\end{centering}
\end{figure}

The shocked ISM produces thermal emission through free-free mechanism (thermal Bremsstrahlung). For completeness we computed this contribution to the total SED. 

We calculated the thermal emission  by integrating the emissivity $\epsilon$ along the shocked ISM, considering the temperature and density profiles given by Eq. (\ref{Tpro}) and (\ref{npro}). The emissivity is given by e.g. (Lang 1999)
\begin{eqnarray}
\epsilon &&\sim 5.4 \times 10^{-39} \frac{n_{i}n_{e}}{\sqrt{T}} g(\nu,T)\exp{-h\nu/k_{\rm B} T}\nonumber\\
&&{\rm erg}\,{\rm s}^{-1}{\rm cm}^{-3}\,{\rm Hz}^{-1}\,{\rm rad}^{-2},
\end{eqnarray}
where $g(\nu,T)$ is the free-free Gaunt factor given by $\sim$  $0.54\ln{\left[ 5 \times 10^{7}\left( T^{3/2}/{\nu}\right) \right] }$; $n_{i}$ and $n_{e}$ are the ion and electron densities respectively, we assumed $n_{i}$ = $n_{e}$ = $n_{\rm fs}$. The emission volume is a spherical wedge of width $x_{\rm c}$ and radius $\Delta$, see Fig. \ref{region}.  In Fig. \ref{Therm} the result of this contribution is shown.  

 For the O4I star the free-free emission is negligible (see Fig. \ref{SEDs}). For the O9I case, at energies around $\sim$ 1 eV this thermal contribution dominates over synchrotron radiation.

\subsection{Absorption}

Gamma rays can be absorbed in the acceleration region immediately after they are created, by photon-photon annihilation. All radiation fields in the acceleration region, thermal and non-thermal,  provide target photons for pair creation.  

The differential opacity for a gamma ray traveling in the direction  $\textbf{e}_{\gamma}$ due to photons of an energy $\epsilon$ in the direction $\textbf{e}_{\rm ph}$ is
\begin{equation}
{\rm d}\tau_{\gamma\gamma} = (1-{\bf e}_{\gamma}{\bf e}_{\rm ph})n_{\epsilon}\sigma_{\gamma\gamma}{\rm d}{\epsilon}{\rm d}{\Omega}{\rm d}l,
\label{dtau}
\end{equation} 
where d$\Omega$ is the solid angle of the emitting surface and $n_{\epsilon}$ is the radiation density. 
The photon annihilation cross-section is (Gould \& Schr\'eder 1967)
\begin{eqnarray}
\sigma_{\gamma\gamma}(\beta) = &&\frac{{\pi}r_{e}^{2}}{2}(1-\beta^{2})\nonumber\\ &&\times\left[2\beta(\beta^{2}-2)+(3-\beta^{4})\ln{\Big(\frac{1+\beta}{1-\beta}\Big)}\right],
\label{cross}
\end{eqnarray}
where $\beta =(1-1/s)^{1/2}$, and $s=E_{\gamma}\epsilon (1-{\bf e}_{\gamma}{\bf e}_{\rm ph})/(m_{e}c^2)^2$. Here, $E_{\gamma}$ and $\epsilon$ are the energies of the gamma ray and the target photon, respectively. The threshold energy  is given by
\begin{equation}
E_{\gamma}\epsilon =\frac{2(m_{e}c^{2})^{2}}{(1-{\bf e}_{\gamma}{\bf e}_{\rm ph})}.
\label{umbral}
\end{equation}

The opacity is 

\begin{equation}
\tau (E_{\gamma}) =  \frac{1}{2}\int_{l} \int_{\epsilon_{\rm th}}^{\epsilon_{\rm max}} \int_{-1}^{u_{\rm max}} (1-u) \, \sigma_{\gamma\gamma}(\beta)
n_{\rm ph}(\epsilon) {\rm d}u {\rm d}\epsilon {\rm d}l .
\end{equation}
Here, $u = \cos \vartheta$, $\vartheta$ is the angle between the momenta of the colliding photons, and $l$ is the photon path across the target radiation field (here $l$ = $\delta$). The target photon fields are those generated within the acceleration region, the IR emission from the heated dust, and the star radiation field at a distance $d$ = $R_{0} - \delta$.

\begin{figure}
\begin{center}
\resizebox{.5\hsize}{!}{\includegraphics{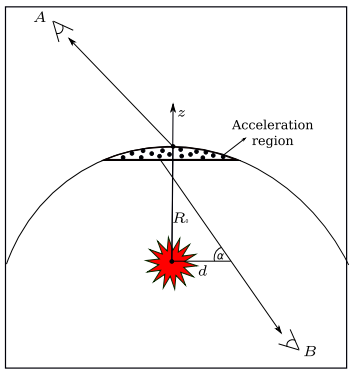}}
\caption{Diagram of the path followed by a gamma ray  traveling from the acceleration region toward  observers placed in front and behind the runaway star (not to scale).}
\label{ext}
\end{center}
\end{figure}

The optical depth is a trajectory integral on which the angular dependence has a very significant effect, and consequently  the absorption depends strongly on the line of sight (e.g. Romero, del Valle, \& Orellana 2010). Photons traveling from the acceleration region toward an observer placed at $A$, see Fig. \ref{ext}, are not absorbed by the stellar photon field. On the other hand, photons traveling toward an observer placed at $B$ strongly interact with the stellar photon field. This interaction depends on the distance $d$ between the star and the photon path, and therefore depends on the angle $\alpha$.

Photons of lower energies  are absorbed by matter through photoionization. The target material can be the shocked ISM and the material along the line of sight corresponding to each particular source. The optical depth  $\tau_{\gamma {\rm H}}$ can be approximated as 
\begin{equation}
{\tau}_{\gamma {\rm H}} \sim N_{\rm H} {\sigma}_{\gamma {\rm N}} (E_{\gamma}). 
\end{equation}
Here $N_{\rm H}$ is the column density corresponding to the target density. The cross section $\sigma_{\gamma {\rm N}}$ was taken from Ryter (1996). For more details see Reynoso, Medina \& Romero (2011) and references therein.

Below we apply the radiative model described so far to the specific case of $\zeta$ Oph. 

\section{Application to $\zeta$ Oph}\label{apli}

The star $\zeta$ Oph (HD 149757) is one of the brightest massive stars in
the northern hemisphere and has been intensively studied. This star has spectral type O9.5V and is a well-known runaway star. It is rapidly rotating with almost break-up velocity, with $v\sin(i)$ $\sim$ $400-500$ km s$^{-1}$ (Walker et al. 1979; Repolust et al. 2004). 

$\zeta$ Oph  bowshock has been observed by the Infrared Astronomical Satellite {\it IRAS} (Van Buren \& McCray 1988, Noriega-Crespo et al. 1997) and more recently by the Wide-field Infrared Survey Explorer {\it WISE}\footnote{$http://wise2.ipac.caltech.edu/docs/release/prelim/$}. The {\it WISE} image shows a very clear, regular structure. This source is quite nearby, located at $\sim$ 222 pc from the Earth (Megier et al. 2009).

\subsection{Bowshock shape}

We adopted a mass loss rate for $\zeta$ Oph of $\sim$ 10$^{-7}$ $M_{\odot}$ yr$^{-1}$ (Fullerton et al. 2006) and terminal wind velocity $V_{\rm w} = 1550$ km s$^{-1}$ (e.g. Hurbig, Oskinova \& Scholler 2011). For the ISM density we adopted a value of $n_{\rm a}$ $\sim$ 10 cm$^{-3}$; since $\zeta$ Oph is embedded in a cirrus cloud region, the density and ambient temperature might be higher than average (e.g. Vidal et al. 2011). These parameters yield $R_{0}$ $\sim$ $0.3$ pc, which agrees well with the upper limit measured by  Peri et al. (2011).  The list of the values of the main parameters adopted in our calculations (we follow Marcolino et al. 2009) are in Table \ref{table}.

\begin{figure}
\includegraphics[trim=0cm 0cm 0cm 0cm, clip=true,width=0.7\linewidth,angle=270]{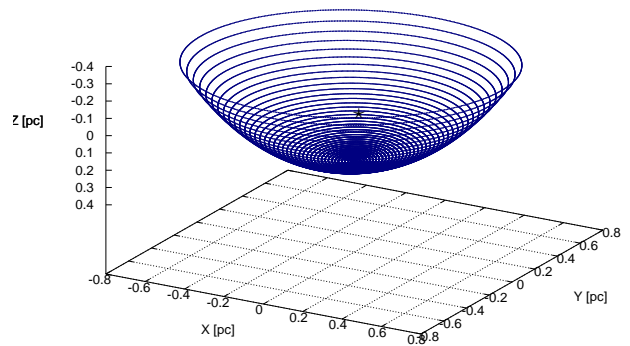}
\caption{$\zeta$ Oph computed bowshock.}
\label{bow3d}
\end{figure}

To compute the bowshock shape we used the analytical method developed by Wilkin (2000). We  assumed that the angular dependence of the wind momentum flux is equatorial ($c_{2} = -1$ and $\lambda$ $= 30\degr$, see Sec. 4.3 of Wilkin 2000). Fig. \ref{bow3d} shows the 3-D computed bowshock shape.

\begin{figure}
\begin{center}
\includegraphics[width=.7\linewidth,angle=0]{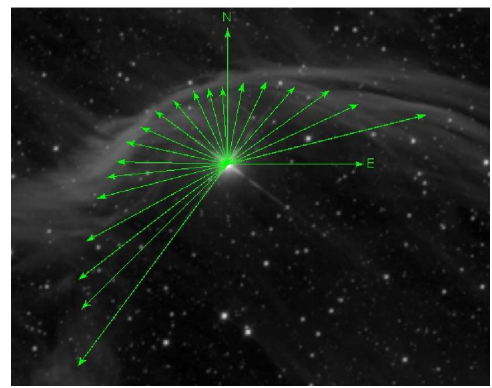}
\includegraphics[trim=0cm 0cm 0cm 0cm, clip=true,width=0.7\linewidth,angle=270]{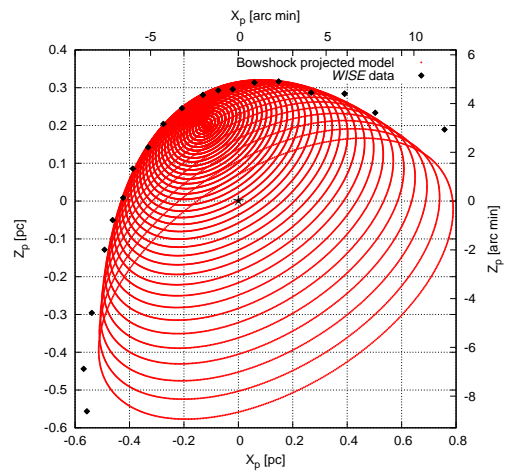}
\caption{Top: image of $\zeta$ Oph from {\it WISE} ($http://wise2.ipac.caltech.edu/docs/release/prelim/$). Dots pointed by arrows correspond to the projection to the 3-D structure. Bottom: $\zeta$ Oph projected bowshock.}
\label{bow2d}
\end{center}
\end{figure}

To compare the 2-D image observed by {\it WISE} we rotated the 3-D coordinate system by three angles. We defined the coordinates $x_{\rm p} \equiv E$ and $z_{\rm p} \equiv N$  (see Fig. \ref{bow2d}) to describe the {\it WISE}  image.  As usual, the origin is placed at the position of the star. From the {\it WISE} image of the bowshock of $\zeta$ Oph it is evident  that the midpoint from the star to the bowshock (see Fig. 4 from Peri et al. 2011) is in the direction of the star velocity.

Figure \ref{bow2d} shows the best fit of $\zeta$ Oph bowshock. 

Our result shows that a simple model can provide a good representation of the observational data, assuming that the IR image is a good tracer of the actual hydrodynamic bowshock shape. The differences between the observed and the intrinsic shape of the bowshock depend on many factors involving radiative transfer, cooling time, dust characteristics, dust distribution, and so on.

\begin{figure}
\begin{center}
\includegraphics[width=0.5\linewidth,angle=270]{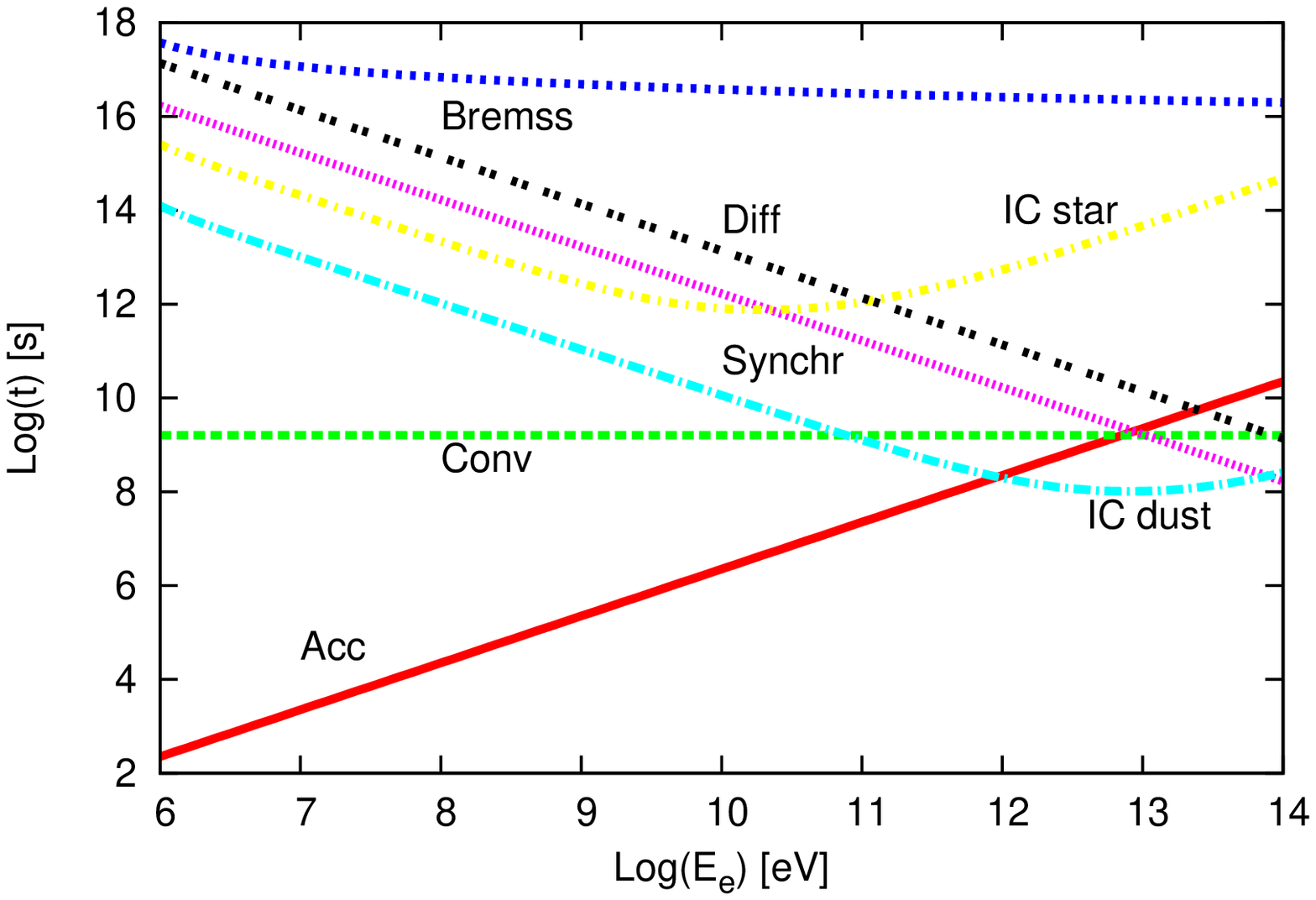}
\includegraphics[width=0.5\linewidth,angle=270]{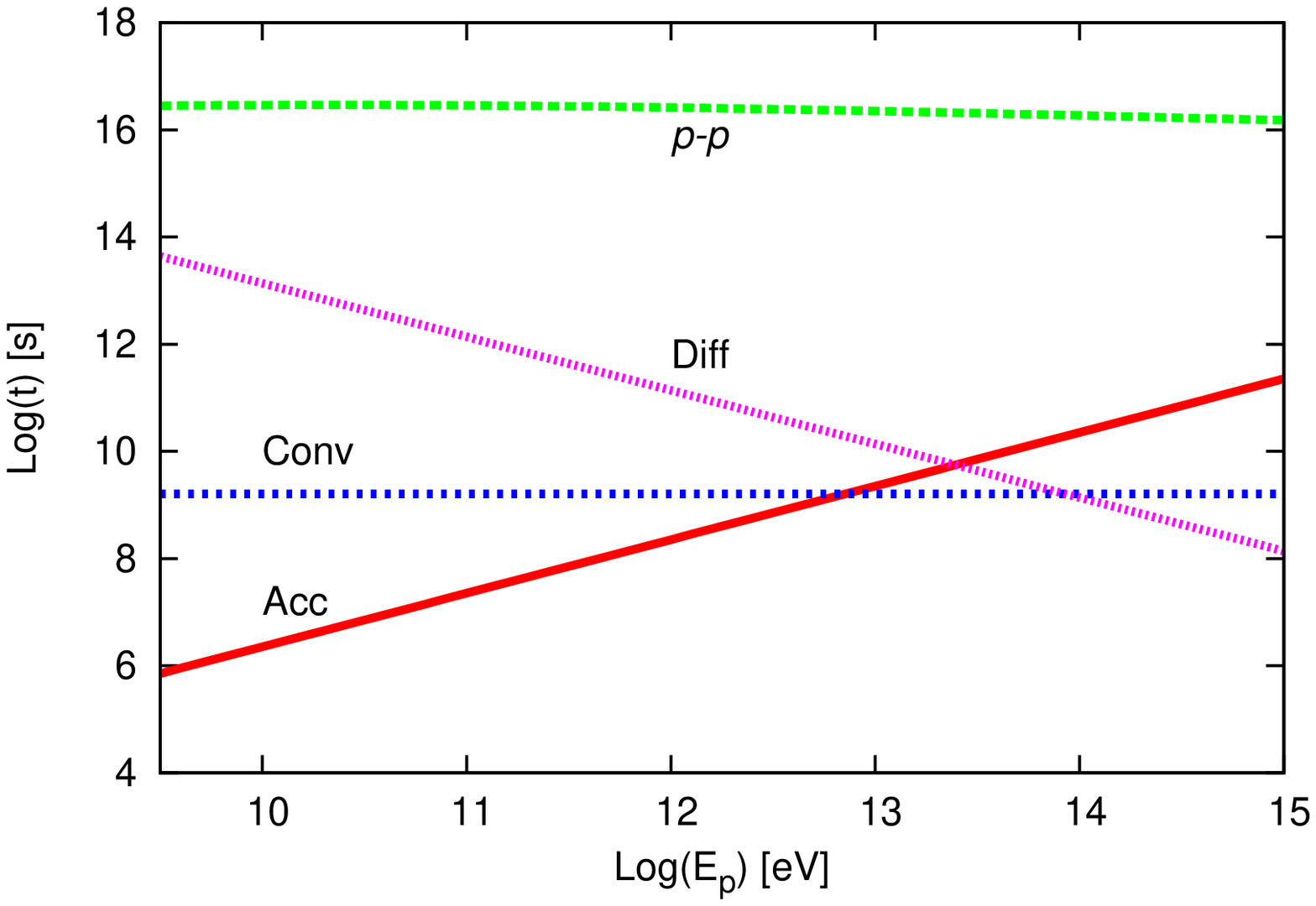}
\caption{Electron and proton losses, acceleration rates, diffusion and convection time scales -- defined in Sec. \ref{diff} --  for $\zeta$ Oph.}
\label{fig:Perdidas}
\end{center}
\end{figure}

\begin{figure}
\begin{center}
\includegraphics[width=0.4\linewidth,angle=270]{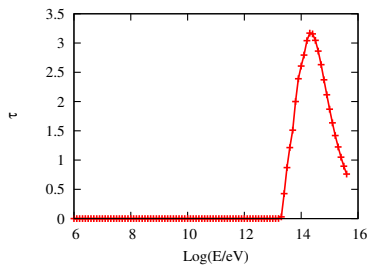}
\end{center}
\caption{Internal opacity.}
\label{TAU}
\end{figure}

\begin{table}
\begin{center}
\caption[]{Parameters for $\zeta$ Oph}
\begin{tabular}{lll}
\hline\noalign{\smallskip}
\multicolumn{2}{l}{Parameter} & value\\
\hline\noalign{\smallskip}
$R_{\rm 0}$ & Standoff radius & 0.3 pc  \\
$\dot{M_{\rm w}}$ & Wind mass loss rate& 10$^{-7}$ M$_{\odot}$ yr$^{-1}$\\
$a$ &Hadron-to-lepton energy ratio & 1 \\
$q_{\rm rel}$ & Content of relativistic particles & 10$\%$   \\
$\alpha$ &Particle injection index & 2\\
$V_{\rm w}$ & Wind velocity & 1.5$\times10^{8}$ cm s$^{-1}$   \\
$L$ & Available power & 5$\times10^{33}$ erg s$^{-1}$\\
$B$ &Magnetic field & 5$\times 10^{-4}$ G  \\
$V_{\star}$ & Star velocity &  30 km s$^{-1}$ \\
$n_{\rm a}$ & ISM number density  & 10 cm$^{-3}$ \\
$T_{\star}$ & Star temperature & 3.2$\times 10^{4}$ K\\
$R_{\star}$ & Star radius & 9 $R_{\odot}$\\
$L_{\star}$ & Star luminosity & $10^{5}$ $L_{\odot}$\\
$T_{\rm IR}$ & Dust temperature & $\sim$ 66  K\\
\hline\\
\end{tabular}	
 \label{table}
\end{center}
\end{table}

\subsection{Spectral energy distribution}

In  Fig. \ref{fig:Perdidas} we  show the radiative losses, the acceleration rates and the diffusion and convection times (see Sec. \ref{diff}) for electrons and protons. The maximum energy is $\sim$ TeV for both species of particles. These values are in accordance with the Hillas criterion (see Sec. \ref{acc}).

The  internal photon-photon optical depth in the bowshock is shown in Fig.\ref{TAU}. It is negligible in the energy ranges of interest. The photoelectrical absorption is also negligible due to the small amount of  material that photons cross on their way to the observer. The external absorption is also  negligible given the relative positions of the bowshock, the star and the observer.

\begin{figure*}
\begin{center}

\includegraphics[width=0.6\linewidth,angle=270]{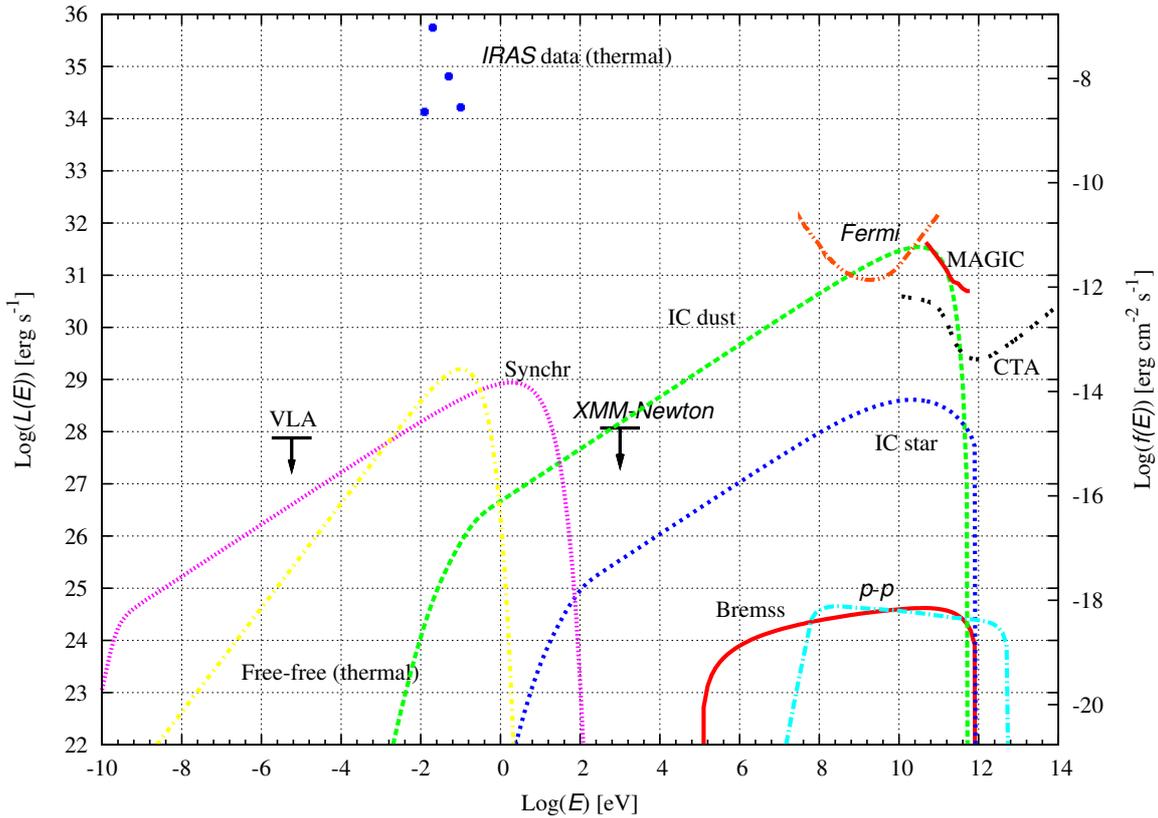}
\caption{Computed SED for $\zeta$ Oph bowshock,  at d $\sim$ 222 pc. The sensitivity for CTA, {\it Fermi}, MAGIC, {\it XMM-Newton} and VLA. {\it IRAS} data are also shown.}
\label{SEDz}
\end{center}
\end{figure*}

Figure \ref{SEDz} shows the computed spectral energy distribution (SED)  for the emission from the bowshock of $\zeta$ Oph, along with the sensitivity  of the gamma-ray detectors CTA (Cherenkov Telescope Array--forthcoming--),  MAGIC and {\it Fermi},  the X-ray satellite  {\it XMM-Newton} (theoretical upper limit from Hasinger et al. 2001), and VLA (upper limit from the NVSS survey -- Condon, Cotton, Greisen et al. 1998 --, angular resolution is given by  Peri et al. 2011).  For completeness the IR {\it IRAS} data are also shown (Van Buren \& MacCray 1988). 

The expected non-thermal luminosity of the source is weak. However, since $\zeta$ Oph is very nearby, the bowshock might be detectable at gamma-ray and X-ray wavelengths through long exposures, under the assumptions we made. We remark that the sensitivity shown in Fig. \ref{SEDz} for MAGIC is for 50 hours of exposure over the source, and for {\it Fermi} it is for one year of integration. For these types of sources an instrument like CTA gives the best chance of detection. CTA might become a unique tool to explore the high-energy radiation produced by runaway massive stars and the population of relativistic particles generated in them. 

\section{Discussion and conclusions}\label{end}

Bowshocks of runaway massive stars are natural candidates for particle acceleration. The different types of massive stars  have different energetics (depending on the wind parameters). The available power for particle acceleration also depends on the distance between the star and the acceleration region. The different available powers produce different non-thermal fluxes. Under the assumptions we made the asymmetries that might arise in runaway bowshocks do not produce a difference in the emitted spectrum. The obtained SEDs depend essentially on the particular assumptions made for the particle acceleration, the magnetic field, and the dust emission.

The emission might be detectable at several wavelengths, provided that the source is close enough and long exposure times are used -- a good candidate is $\zeta$ Oph --. The synchrotron  emission expected at radio wavelengths might be detectable, as in the case of BD +43$\degr$ 3654. The undetectability can establish constraints on  parameters such as the magnetic field in the shocked wind.  Stellar bowshocks might also be detectable at X-ray wavelengths, although no runaway bowshock has been  observed at these energies so far. Finally, a system like $\zeta$ Oph might be detectable at $\gamma$-rays by the future ground-based detector CTA,  as well as by the {\it Fermi} satellite. The energy range between 1 GeV-1 TeV offers the best prospects for the study of runaway stars as non-thermal emitters. Our work suggests that bowshocks of runaway stars might constitute a new class of high-energy sources to be explored in the near future.

\begin{acknowledgements}
We are grateful to Paula Benaglia, Anabella Araudo, Florencia Vieyro and Cintia Peri for insightful discussions. This publication  used  data products from the Wide-field Infrared Survey Explorer, which is a joint project of the University of California, Los Angeles, and the Jet Propulsion Laboratory/California Institute of Technology, funded by the National Aeronautics and Space Administration. This work is supported  by PIP 0078 (CONICET) and PICT 2007-00848, Pr\'estamo BID (ANPCyT). G.E.R. received additional support from the Spanish Ministerio de 
Inovaci\'on y Tecnolog\'{\i}a  under grant AYA 2010-21782-c03-01. 
\end{acknowledgements}

{}
\end{document}